\documentclass[manuscript]{aastex}	
\usepackage{epsfig}			
\usepackage{graphicx,color}		
\usepackage{amssymb}			
\usepackage{color}			
\usepackage{url}			
\usepackage{amsmath}			
\usepackage{rotating}			
\usepackage{float}			
\usepackage{textcomp}			
\usepackage{psfig}
\usepackage{epstopdf}
\usepackage{dcolumn}
\usepackage{times}
\usepackage{tabularx}
\usepackage[english]{babel}
\usepackage{subcaption}
\usepackage{ragged2e}

\shorttitle{Solar Active Longitude Revisited}
\shortauthors{S. Mandal et al.}

\begin{document}

\title{Solar Active Longitudes From Kodaikanal White-light Digitized Data}

\author{Sudip Mandal$^{1}$,
Subhamoy Chatterjee$^{1}$,
Dipankar Banerjee$^{1,2}$
}
 
\affil{$^{1}$Indian Institute of Astrophysics, Koramangala, Bangalore 560034, India. e-mail: {\color{blue}{sudip@iiap.res.in}}\\
$^{2}$ Center of Excellence in Space Sciences India, IISER Kolkata, Mohanpur 741246, West Bengal, India  \\
}
  \justify
  
\begin{abstract}
The study of solar active longitudes has generated a great interest in the recent years. In this work we have used an unique continuous sunspot data series obtained from Kodaikanal observatory and revisited the problem. Analysis of the data shows a persistent presence of the active longitude during the whole 90 years of data duration. We compare two well studied analysis methods and presented their respective results. The separation between the two most active longitudes is found be roughly 180\textdegree~ for majority of time. Additionally, we also find a comparatively weaker presence of separations at 90\textdegree~ and 270\textdegree~. Migration pattern of these active longitudes as revealed from our data is found to be consistent with the solar differential rotation curve. We also study the periodicities in the active longitudes and found two dominant periods of $\approx$1.3 years and $\approx$2.2 years. These periods, also found in other solar proxies, indicate their relation with the global solar dynamo mechanism.

\end{abstract}

\section{Introduction}
Sunspots are the prominent features, on the solar photosphere, visible in white light. Sunspots show a preferred latitudinal dependence which moves from higher latitude to towards the equator with the progress of the 11-year sunspot cycle. This migration pattern of the activity zone is known as `butterfly diagram'. Similar to the preferred latitudinal belt, solar active longitudes refer to the longitudinal locations with higher activity compared to the rest of the Sun. Active longitudes in the past has been studied for solar like stars  \citep{1998A&A...336L..25B,2000A&A...358..624R}. There have been quite a few studies on the active longitudes, from the observational data as well as from the numerical simulations  \citep[see, for a complete review]{lrsp-2005-8}. 

One of the earliest works on the solar active longitudes had been published from Kodaikanal observatory by \citet{1932MNRAS..93..150C}. Other notable works were by \citet{1939POMic...7..127L,1961Obs....81..205L}. In recent times, \citet{1983SoPh...87...23B} (and references therein) showed a strong correlation between the active longitude and the high speed solar wind. Most of the previous works showed the existence of active longitude on smaller time scales of 10-15 Carrington rotations. Using Greenwich sunspot data, \citet{2003A&A...405.1121B} reported the presence of two active longitudinal zones which are persistent for more than 120 years. These two zones alter their activity periodically between themselves. Apart from this `flip-flop' like behavior, these authors have also shown that the active longitudes move as a rigid structure i.e the separation between the two active longitudes is roughly a constant value of 180\textdegree~. Migration of the active longitude in the sunspot, is studied by \citet{2005A&A...441..347U} where the authors have found that the migration pattern is governed by the solar differential rotation. In this work they have invoked the presence of a weak non-axisymmetric component in the solar dynamo theory in order to explain the observed longitudinal patterns. There are some criticism of these works too. \citet{2005A&A...429.1093P} have shown that some of the above quoted results may be an artefact of the methods used to derive them.

Active longitudes have also been discovered in other solar proxies. Solar flares, specially proton flares are found be associated closely with the active longitude locations \citep{1969SoPh....6..104B}. \citet{2000JGR...105.2315N} found existence of preferred longitudes in the near-Earth and near-Venus solar wind data. Analyzing the X-ray flares observed with NOAA/GOES satellite, \citet{Zhang2007970} have shown the presence of active longitudes as well as their migration with time. However the differential rotation parameters obtained with the X-ray flares were found to be different from those obtained by using the sunspots in \citet{2005A&A...441..347U}. Using a combination of Debrecen sunspot data and RHESSI data, \citet{2016ApJ...818..127G} established a probable dependence of the flare occurrence with the active longitudes.

In this paper we use the Kodaikanal white-light digitized data for the first time and revisited the active longitude problem with multiple analysis approaches. In Section~\ref{data_description} we give a brief description of the Kodaikanal data which we have analyzed,  using two recognized methods used in the literature. In Section~\ref{u_method} we describe the rectangular grid method and the subsequent results from that. We also study the effect of the sunspot size distribution in this method as shown in Section~\ref{area_thresholding}. The other method called as `bolometric method' has been described in Section~\ref{b_method}. Periodicities and the migration pattern in the active longitudes is described in Section~\ref{period} and in section~\ref{theory_mig} respectively, followed by the `summary and the discussion' in the end.
\begin{figure*}[!htbp]
\centering
\includegraphics[width=0.85\textwidth]{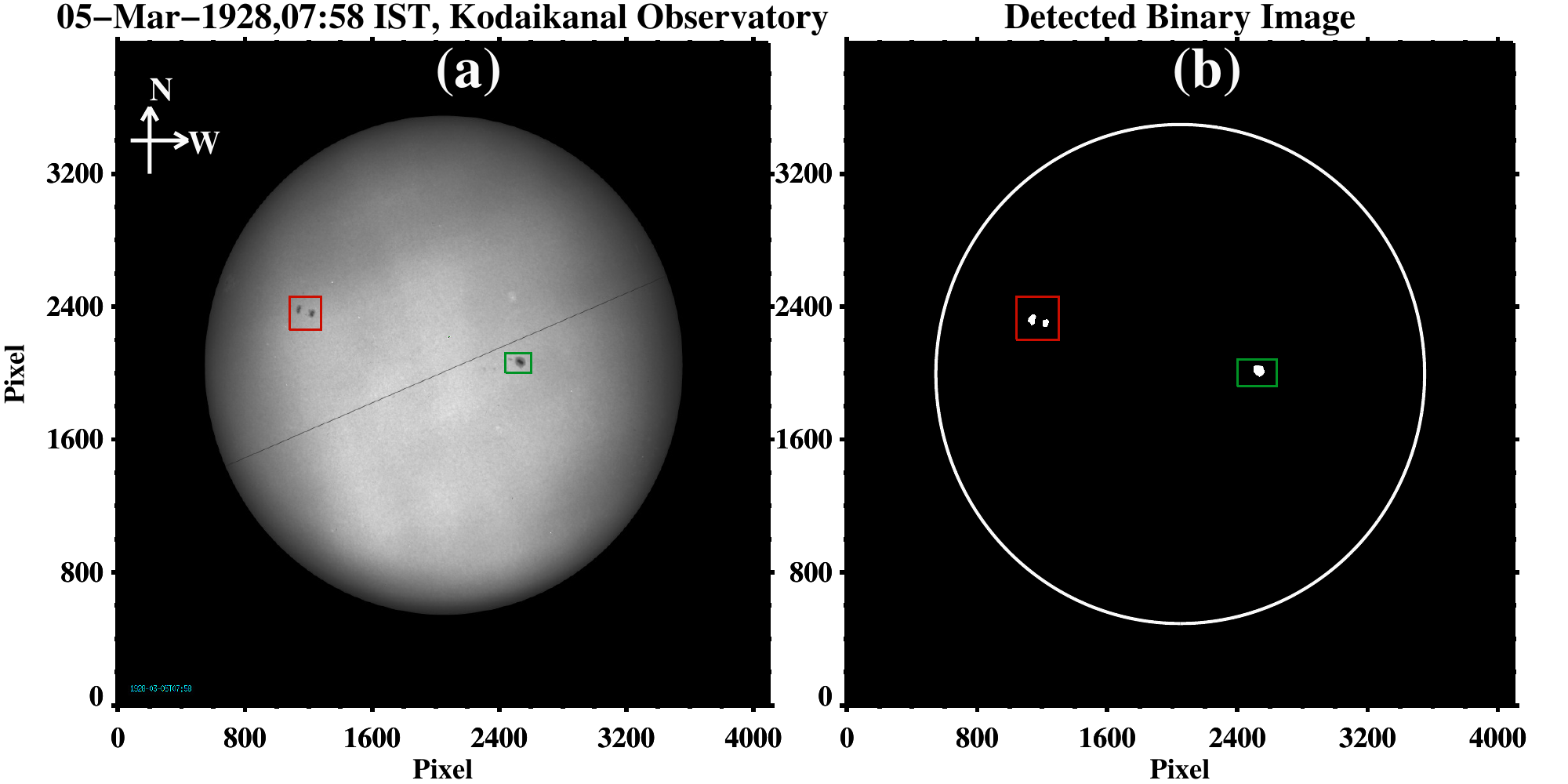}

\caption{Panel (a) shows a representative image of the Kodaikanal white-light digitaized data. Observed sunspots on the image has been highlighted by two rectangular boxes. The binary image of these detected sunspots is shown in Panel (b). The solar limb has been overplotted for better visualization.} 

\label{kodai_context}
\end{figure*}
\section{Kodaikanal Data Description} \label{data_description}

 We have used the white-light digitized sunspot data from the Kodaikanal observatory, India. The data period covers more that 90 years, starting from 1921 to 2011. Original solar images were stored on photographic plates and films and were preserved carefully in paper evelopes. These images has been recently digitized (in 4k$\times$4k format) by \citet{2013A&A...550A..19R}. Using a modified STARA algorithm \citep[see, for details]{2013A&A...550A..19R} on this digitized data, sunspot parameters like area, longitude, latitude etc have been extracted by \citet{2016arXiv160804665M} (henceforth paper-~$\mathrm{I}$). Apart from comparing the Kodaikanal data with data from other observatories, in paper-~$\mathrm{I}$ we have also discussed about different distributions in sunspot sizes in latitude as well as in longitude. While detecting the sunspots, images of the detected sunspots were also saved in a binary format. Panel (a) in Figure~\ref{kodai_context} shows a representative full disc digitized white-light data. Two rectangular boxes highlight the sunspots in two hemispheres present on that day. The binary image containing these detected sunspots is shown in panel (b) of Figure~\ref{kodai_context}.
\subsection{Generation of Carrington Maps}
\begin{figure*}[!htbp]
\centering
\includegraphics[width=0.77\textwidth]{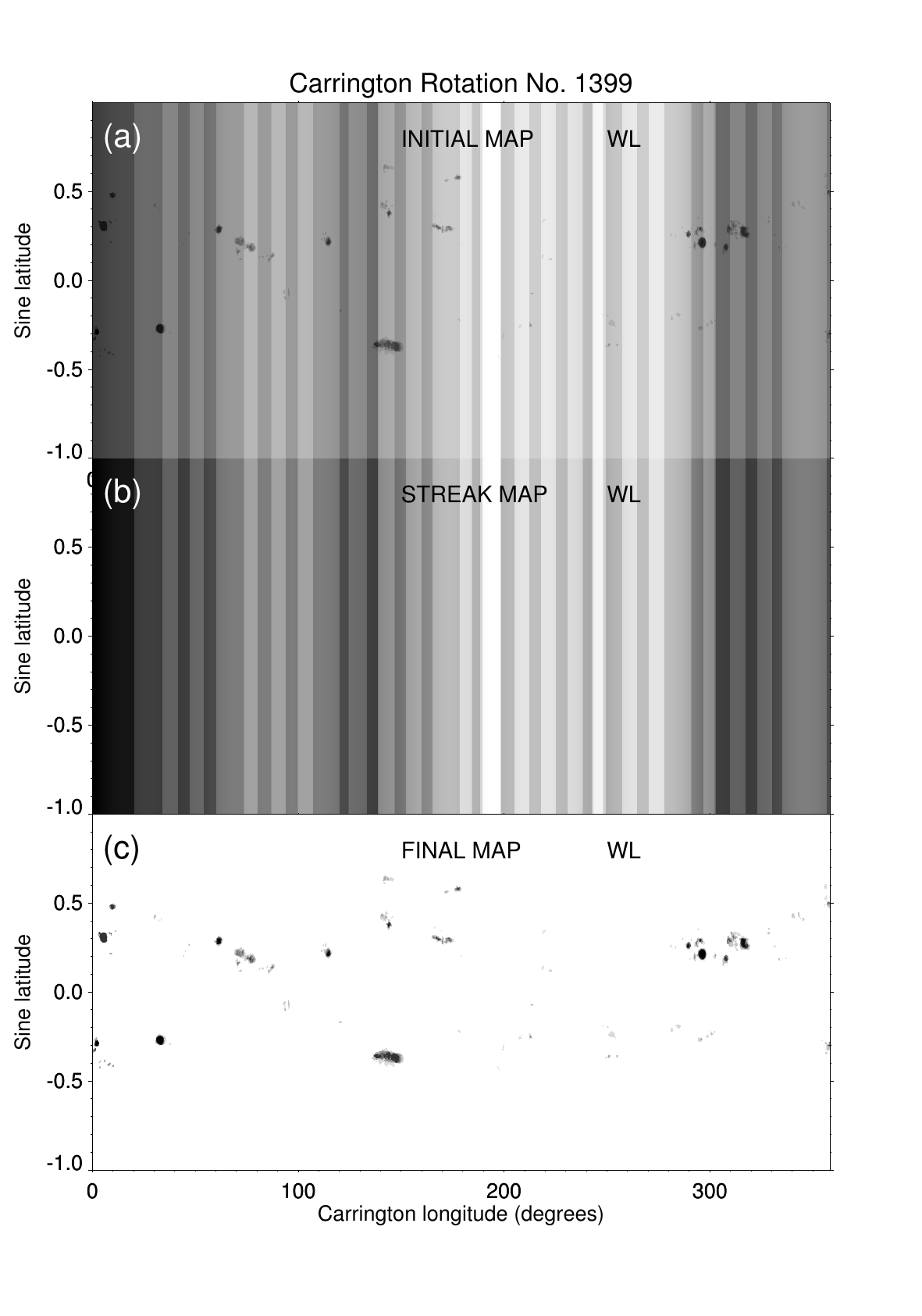}

\caption{Different steps of producing a Carrington map from Kodaikanal white-light data are shown above for a representative Carrington rotation number 1399. Panel (a) shows the original map produced from the binary images. The streak map shown in panel (b), has been used to create the final map (panel (c)) from the original map. } 

\label{carr_context}
\end{figure*}

  Carrington maps are the Mercator projected synoptic charts of the spherical Sun in Carrington reference frame \citep{1998ASPC..140..155H}. We have used the daily detected sunspot images (as shown in panel (b) of Figure~\ref{kodai_context}) to construct the Carrington maps. A longitude band of 60\textdegree~ (-30\textdegree~ to +30\textdegree~ in heliographic coordinate) is selected for each image to construct these maps (following \citealp{2011ApJ...730...51S}). This involves stretching, B$_{\circ}$ angle correction (B$_{0}$ angle defines the tilt of the solar north rotational axis towards the observer. It can also be interpreted as the heliographic latitude of the observer or the center point of the solar disc), shift in the Carrington grid and additions. One Carrington map has been constructed considering a full 360\textdegree~ rotation of the Sun in 27.2753 days. In order to correct for the overlaps, we have used the `streak map'\citep{2011ApJ...730...51S} for every individual Carrington map and divided the original maps with them. Data gaps occur as black longitude bands in these maps. The whole procedure is shown in various panels of Figure~\ref{carr_context}. Here we must emphasize the fact that in our Kodaikanal data we have some missing days (the complete list of missing days has been published with Paper-~$\mathrm{I}$). In order to increase the confidence on the obtained results, we have not considered any Carrington map in our analysis which has one or more missing days.

\section{Data Analysis}

 We use the generated Carrington maps for our further analysis. Two different methods, `rectangular grid' method and the `bolometric curve' method, are used as described in the following sub-sections. One should note that a possible drift of the active longitudes, due to the differential rotation of the sun, is not considered here but later analyzed in section 4.1.

\subsection{Using Rectangular Grid} \label{u_method}

First we follow the `rectangular grid' method (following \citealp{2003A&A...405.1121B}) where a full Carrington map has been divided in 18 rectangular strips, each of 20\textdegree~ longitudinal width. We then compute a quantity `Weight' ($\mathrm{W}$) defined as 

\begin{equation}
 \mathrm{W}_{i}=\frac{\mathrm{S}_{i}}{\sum_{j=1}^{18} \mathrm{S}_{j}}
\end{equation}
where $\mathrm{S}_{i}$ is the total sunspot area in the $\mathrm{i}^{th}$ bin. We note down the longitudes of the highest and the second highest active bins and calculate the separation between them (afterwards referred as `longitude separation'). We impose a minimum of 20\% peak ratio between the second highest and highest peak in order to avoid any sporadic detection.

\begin{figure*}[!htbp]
\centering
\includegraphics[width=0.85\textwidth]{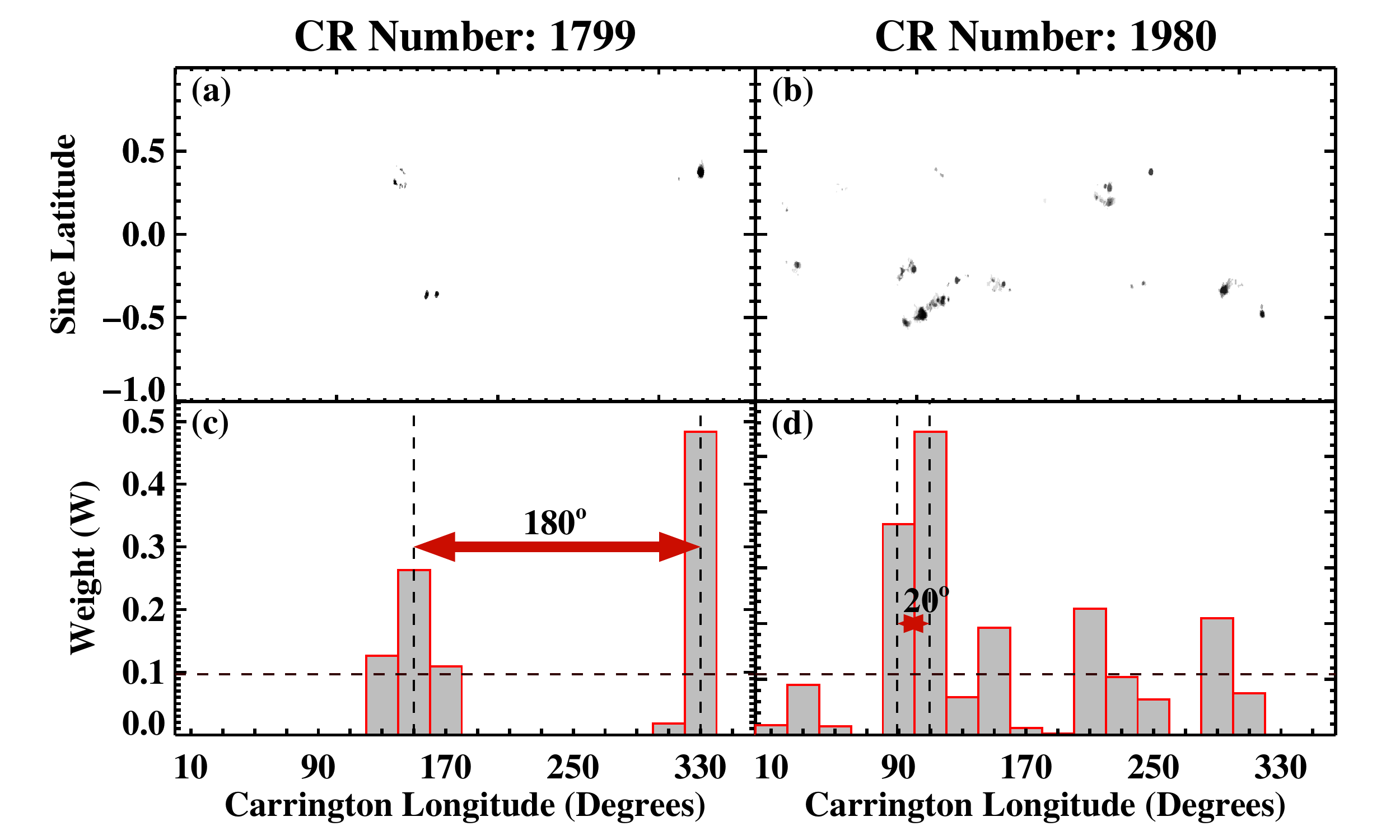}

\caption{Panels (a-b) represent two representative Carrington maps for rotation number 1799 and 1980 respectively. Panels (c-d) show the weighted value ($\mathrm{W}$) in each longitudinal bins for the two maps. Two horizontal dashed lines represent the cut-off value mentioned in the text. The difference between the two highest peaks in each case, are also highlighted in the respective panels.} 

\label{u_bar}
\end{figure*}
\begin{figure*}[!htbp]
\centering
\includegraphics[width=0.90\textwidth]{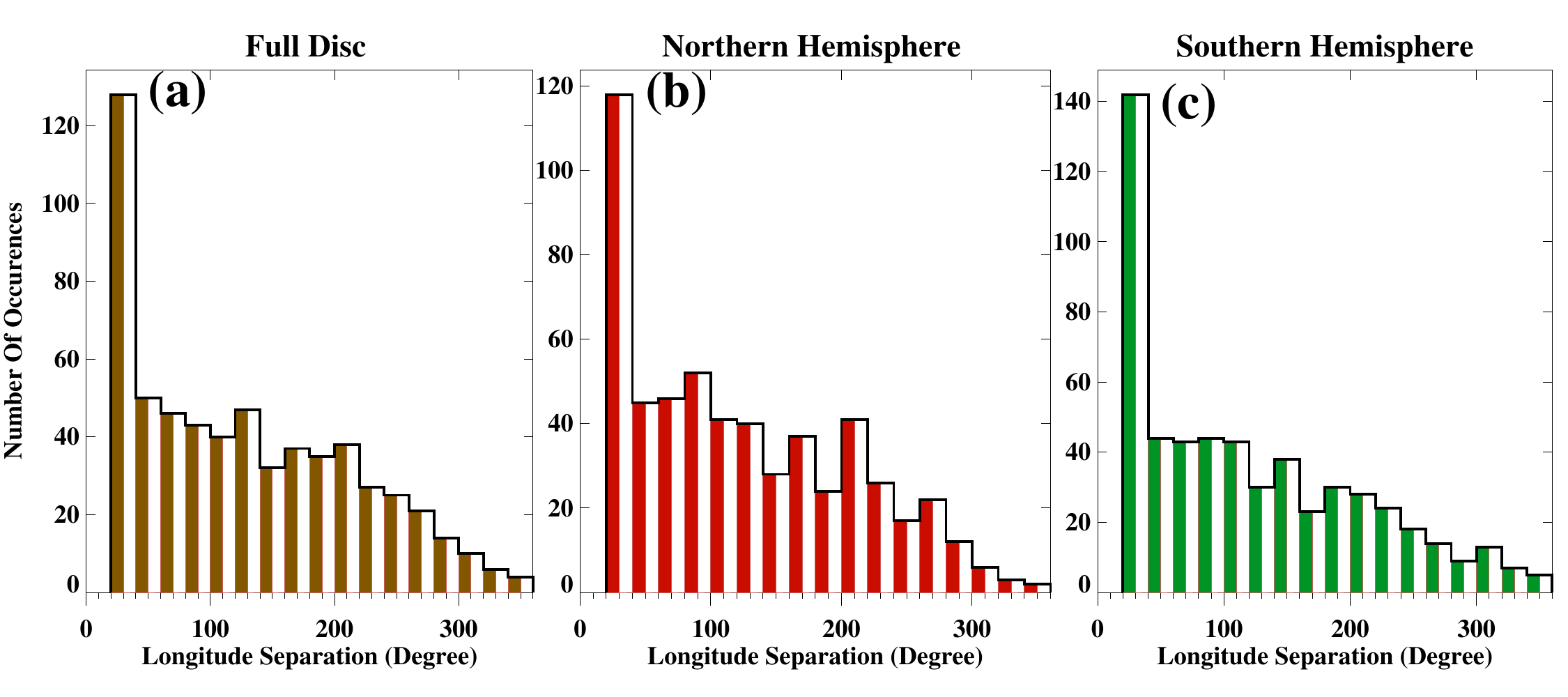}

\caption{Figure shows histograms of the longitudinal difference values between the two most active bins for the full disc, northern and southern hemispheres.} 

\label{u_histograms}
\end{figure*}

 Two such representative Carrington maps from Kodaikanal white-light data archive is shown in panel (a-b) of Figure~\ref{u_bar} and their corresponding barplots in panels (c-d). For the two representative cases shown in the figure, we notice that for CR number 1799 the longitude separation is 180\textdegree~ whereas for CR number 1980 the difference is 20\textdegree~. We compute such longitude separations for each and every Carrington map for the whole hemisphere (referred as `full disc' henceforth) and for individual northern and southern hemispheres. Histograms, constructed using these separation values for each of the three mentioned cases, are shown in different panels of Figure~\ref{u_histograms}. In all three cases (panels (a-c) in Figure~\ref{u_histograms}) we see that the maximum occurrence is for the 20\textdegree~ separation. Apart from that, we also notice peaks at $\sim$120\textdegree~ and at $\sim$200\textdegree~ for the full disc case whereas these peaks shifts a little bit for the northern and southern hemispheric cases. Apart from these mentioned peaks, for the northern and southern hemispheres, we also see weak bumps in the histograms at $\sim$160\textdegree~ and $\sim$270\textdegree~. In an earlier work using Greenwich data, \citet{2003A&A...405.1121B} had reported a phase difference of 0.5 (180\textdegree~ in terms of longitude) between the two most active longitude bins.

\begin{figure*}[!htbp]
\centering
\includegraphics[width=.90\textwidth]{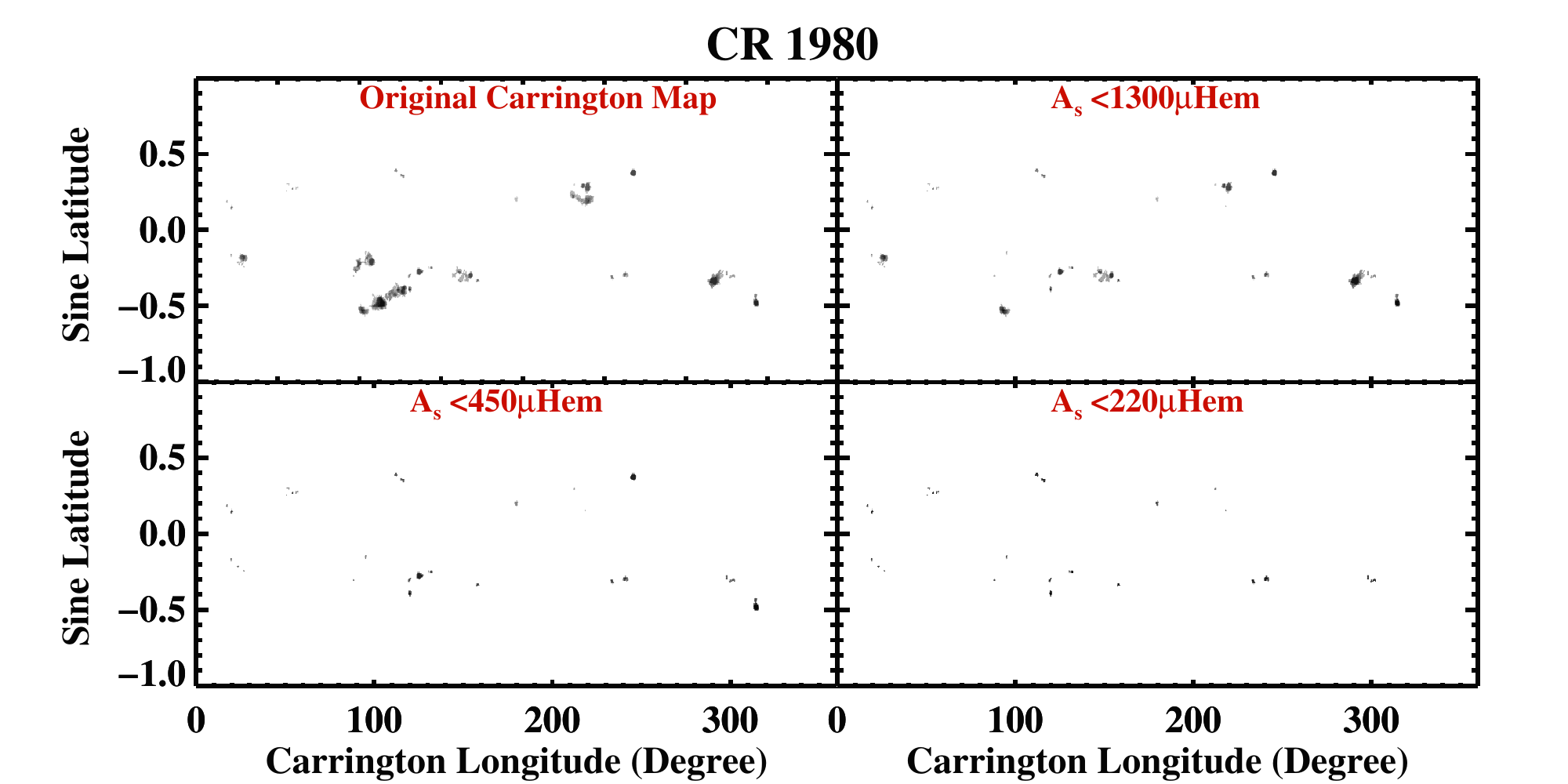}

\caption{ Representation of area thresholding for the CR 1980. Different panels are showing the area thresholded maps for different area values (A$_s$) mentioned in the corresponding panels.} 

\label{t_demo}
\end{figure*}

The reason of this high number of occurrence, of the longitude separations, at $\sim$20\textdegree~ is probably related to the longitudinal extent of the sunspots compared to the chosen bin width of 20\textdegree~. To be specific, there are frequent cases when the largest sunspot or sunspot groups get shared by the two consecutive longitude bins. Now these occurrences are on statistical basis and thus increasing the bin size only shifts the highest peak to the chosen bin value (e.g for a 40\textdegree~ bin size we find the maximum at 40\textdegree~). Since this effect is related with the sunspot sizes, we thus use area thresholding on the sunspots and note down the longitude separations as described in the next section.
\subsubsection{Area thresholding and Active Longitude} \label{area_thresholding}

 We now use the area thresholding method on the sunspots found in every Carrington maps. One such illustrative example is shown in Figure~\ref{t_demo}. We again use the CR 1980 for demonstration as it has sunspots of various sizes. Different panels in Figure~\ref{t_demo} show the Carrington map before and after doing different area thresholdings.
\begin{figure*}[!htbp]
\centering
\includegraphics[width=.48\textwidth]{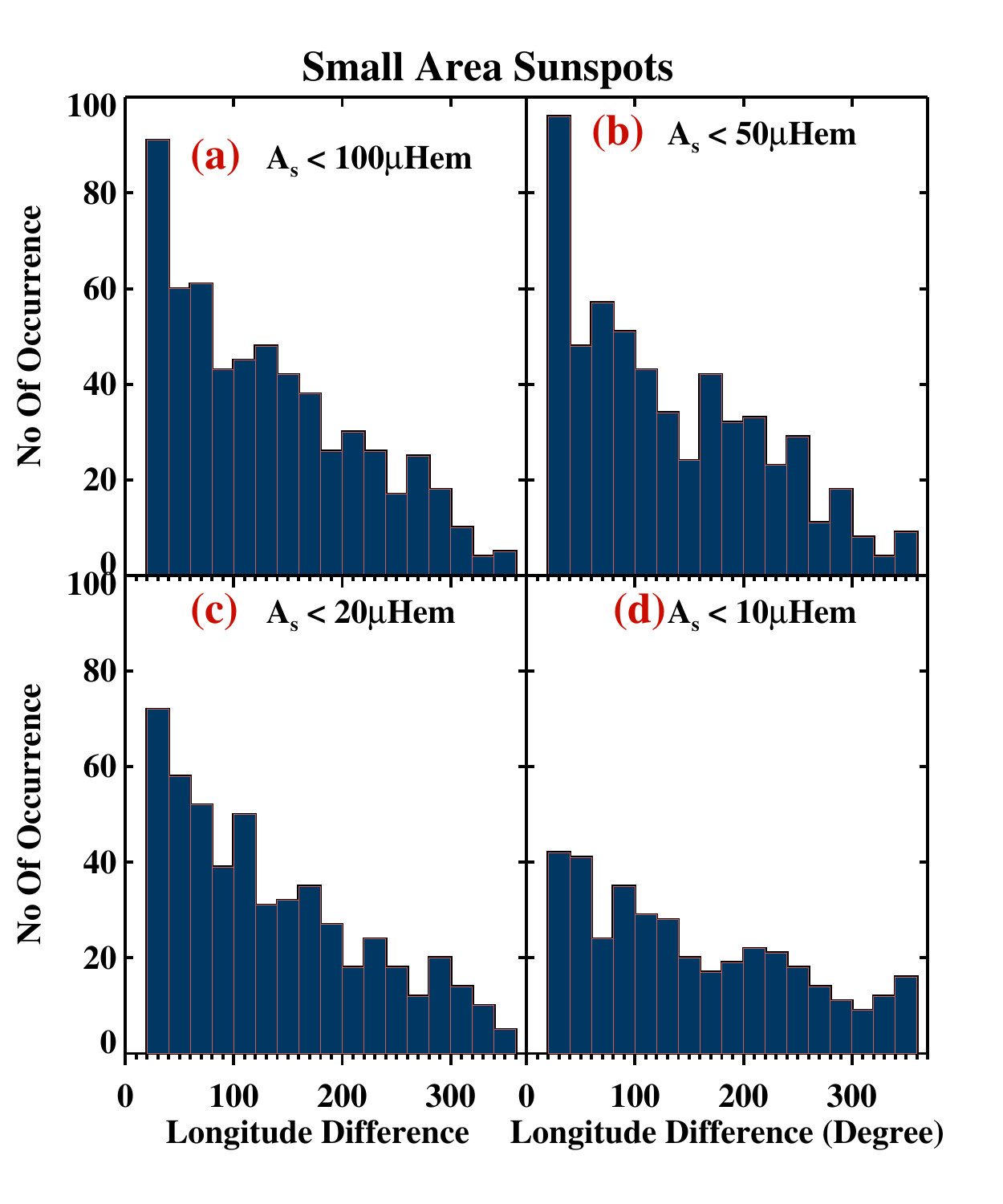}
\includegraphics[width=0.48\textwidth]{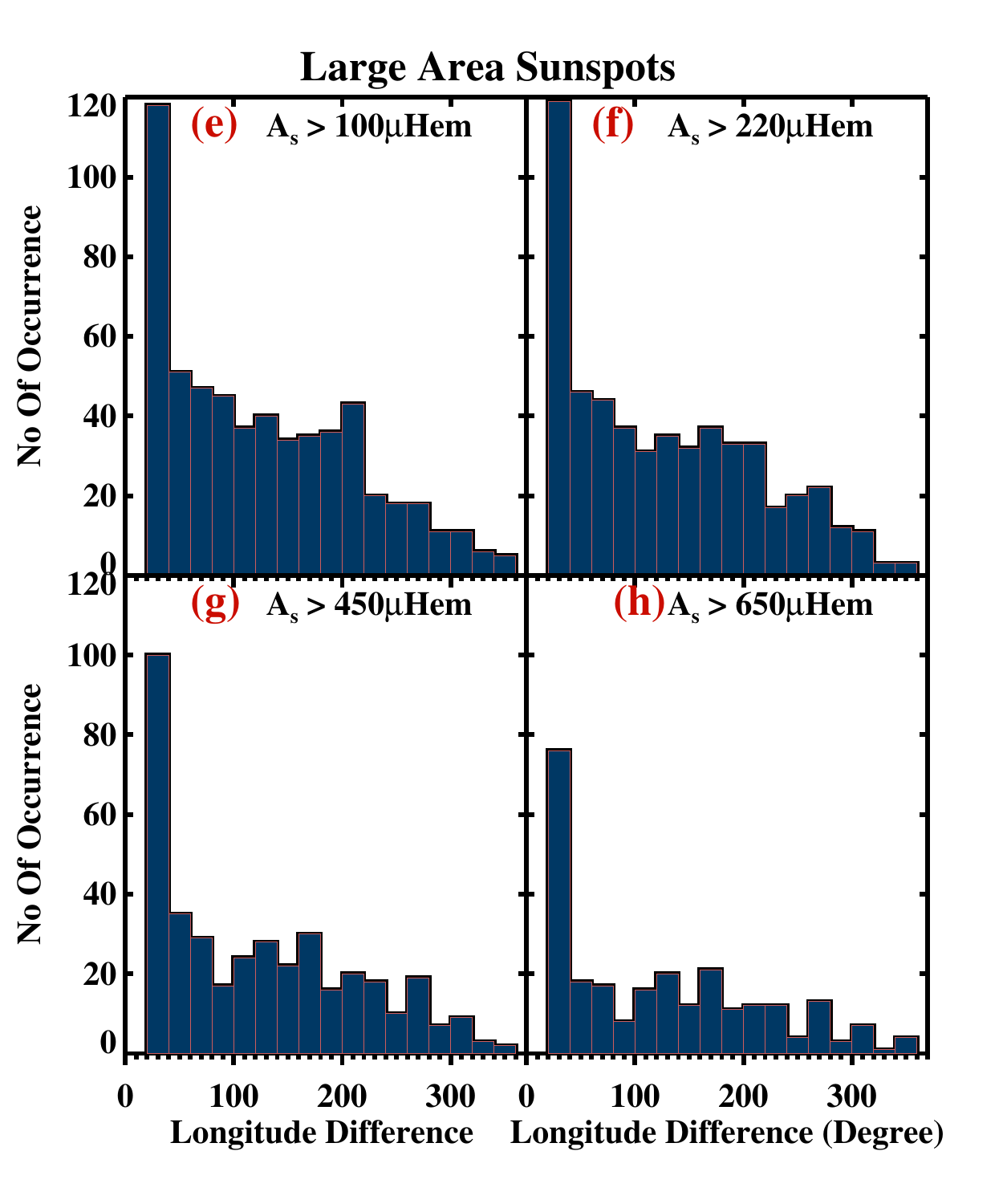}

\caption{ Panels (a-d) show the longitude difference histograms for different area thresholding for the small sunspot areas. Panels (e-h) are similar as before but for larger sunspot area thresholding.} 

\label{area_th}
\end{figure*}

 After performing the area thresholding, we follow the same procedure as described in earlier sub-section to find the longitude separation between the two most active bins.

 Following the extracted sunspots area distribution from the Kodaikanal white-light data (as shown in Fig.7 in \citealp{2016arXiv160804665M}), we chose two kind of area thresholding values. First set of values correspond to sunspots with smaller sizes. These values range from 10 $\mu$Hem to 100$\mu$Hem as shown in panels (a-d) in Figure~\ref{area_th}. From this figure, we immediately notice that the height of the histogram peak at $\sim$20\textdegree~ decreases progressively with the decrease of sunspot sizes. This is explained by the fact that as sunspot sizes go down, the probability of a sunspot being shared by two longitude bins also reduces. This result in a lower peak at $\sim$20\textdegree~. Also, in every case, we notice prominent peaks in the histograms at $\sim$90\textdegree~ and at $\sim$180\textdegree~ separations.

We next investigate the longitude separation for the sunspots with larger sizes. In this case the thresholding values range from 100$\mu$Hem to 650$\mu$Hem. In Figure~\ref{area_th}~(e-h) we show the histograms of the longitude difference for these different thresholds. We see two noticeable differences in this case compared to the former one of small sunspot sizes. Firstly, the peak height of the histograms $\sim$20\textdegree~ increases (relative to the other peak heights) as we move towards the larger sunspots which is expected due to the reason we discussed earlier. Along with that we notice that there is no peak near 90\textdegree~ as found earlier but a peak $\sim$180\textdegree~ is still present along with other new peaks. Here we should highlight the fact that with  higher sunspot area thresholding, the statistics become poor and the peaks become less significant statistically.

\subsection{Using Bolometric Curve} \label{b_method}

 From our previous analysis we saw that the discreteness introduced due to the longitude bins has a definitive effect on the calculated longitude separation. Therefore, we explore the other method, the 'bolometric curve' method \citep{2003A&A...405.1121B}, which produces a smooth curve as described below.

\begin{figure*}[!htbp]
\centering
\includegraphics[trim = 0mm 0mm 0mm 1mm, clip,width=.95\textwidth]{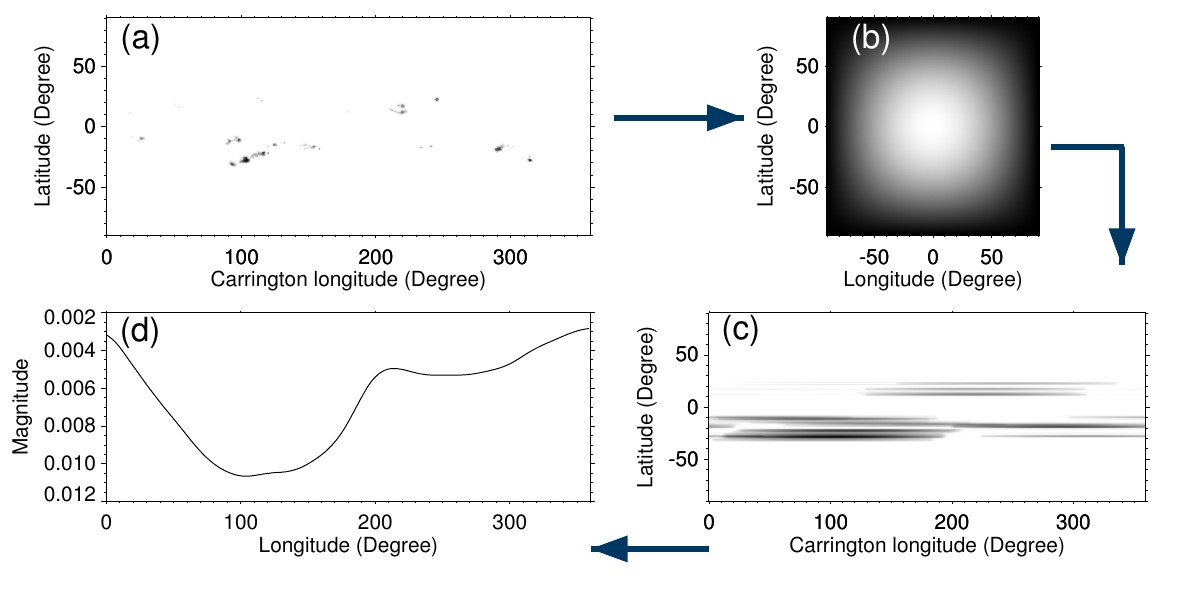}

\caption{Steps to generate bolometric magnitude curve starting from a white-light Carrington map as shown in  (a).;  (b) Limb darkening profile in latitude and longitude; (c) Modified Carrington map after shifting (b), multiplying and adding with (a) along longitude; d) Bolometric magnitude curve generated after adding (c) along latitude normalised by the total of (b). Arrows represent the sequence of bolometric magnitude curve generation.} 

  \label{fig:bolometric}
\end{figure*}

In order to generate the smooth bolometric profile, we first invert the intensities of the Carrington maps to make the white background black with sunspot as a bright feature. Next, the map is stretched to convert sine latitude into latitude (Figure~\ref{fig:bolometric}a). We then generate a limb darkening profile with the expression, profile = $0 .3 + 0.7\mu$ in latitude and longitude where $\mu$ is the cosine of heliocentric angle (Figure~\ref{fig:bolometric}b). We then create an intermediate map by shifting the limb darkening profile, multiplying and adding with the intensity inverted Carrington map along every longitude (Figure~\ref{fig:bolometric}c).

 In the end, this intermediate map is added along latitude for each longitude to generate a factor, called $f$. The $f$ curve is converted to a bolometric magnitude curve ($m$) (Figure~\ref{fig:bolometric}d) using the expression defined as
\begin{equation}
 \mathrm{m} = -2.5\log[\frac{(1-f){T_{ph}}^4+f\times{T_{sp}}^4}{{T_{ph}}^4}]
\label{b_equation}
\end{equation}
 where bright surface temperature $T_{ph}$= 5750~$\mathrm{K}$ and sunspot temperature $T_{sp}$= 4000~$\mathrm{K}$.

\begin{figure*}[!htbp]
\centering
\includegraphics[width=.99\textwidth]{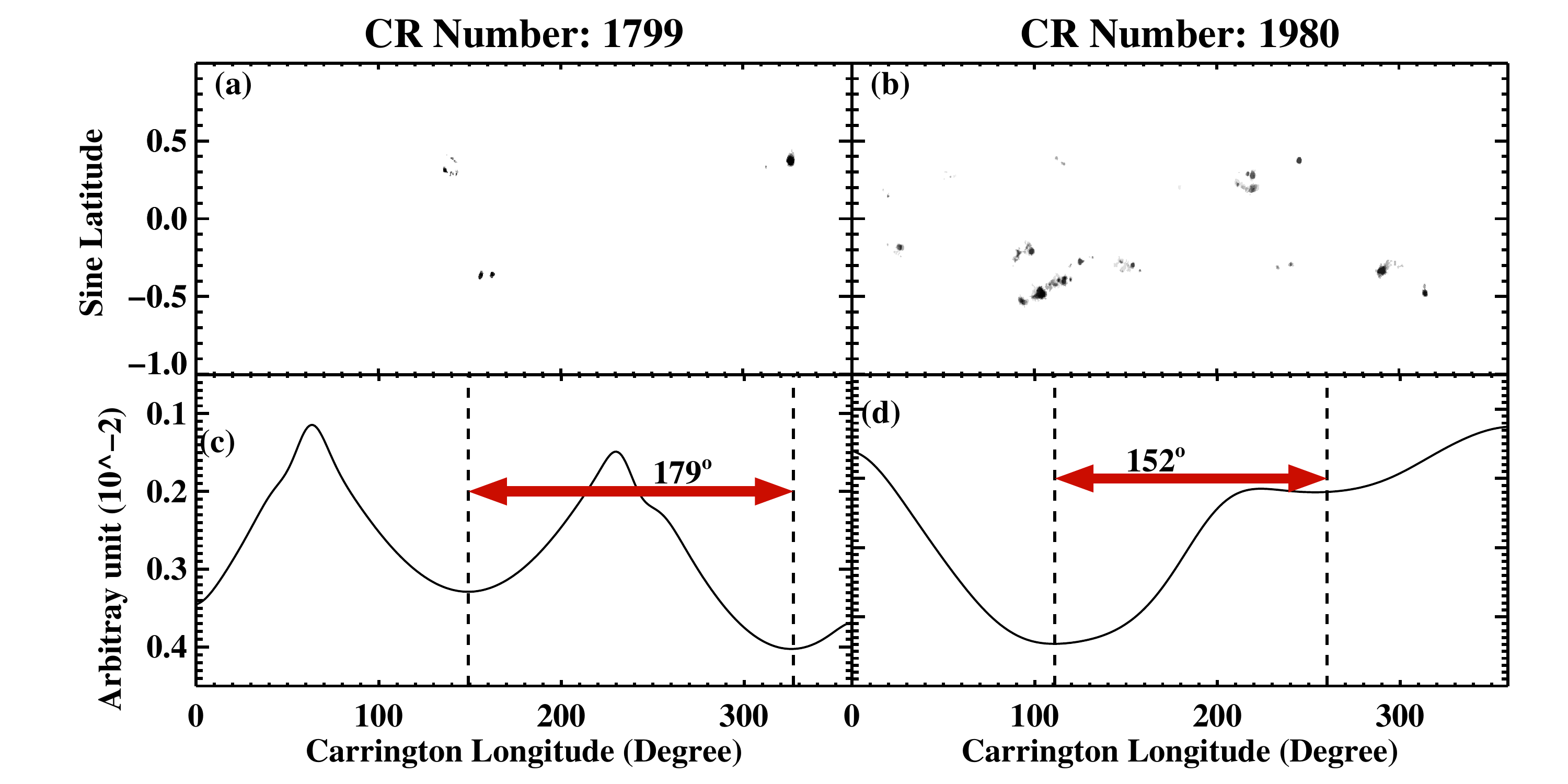}

\caption{Figure shows the bolometric curves (panel (c-d)) for the CR 1799 and CR 1980 respectively (Panels (a-b)). Difference between the longitudes are written in respective panels.} 

\label{b_image}
\end{figure*}

 Figure~\ref{b_image} shows the two representative plots of this bolometric method. We chose the same two Carrington maps as shown in Figure~\ref{u_bar} for easy comparison between the two methods. Now we see that for the CR 1799, the bolometric curve basically traces the active bars (as shown in panel (c) of Figure~\ref{u_bar}) i.e the minimum of the bolometric curve represent the locations of maximum spot concentrations. We notice that the separation in this case (179\textdegree~) equals to the separation obtained previously (180\textdegree~) but for the CR 1980, the difference in this case is $\sim$150\textdegree~ whereas previously obtained value was 20\textdegree~. This is because the bolometric curve takes into account of close spot concentrations contradictory to the fixed longitudinal bins as defined in the rectangular grid method. In principle if we smooth out the peaks shown in Panel (d) of Figure~\ref{u_bar}, we should then arrive with a similar curve as of the bolometric one but the amount of smoothing is  subjective and may be different for different Carrington maps. However in the case of bolometric method, we must emphasize that the bolometric curve has been generated using a fixed prescription (Equation~\ref{b_equation}) and thus it is free from any subjectivity issue.%
\begin{figure*}[!htbp]
\centering
\includegraphics[width=0.75\textwidth]{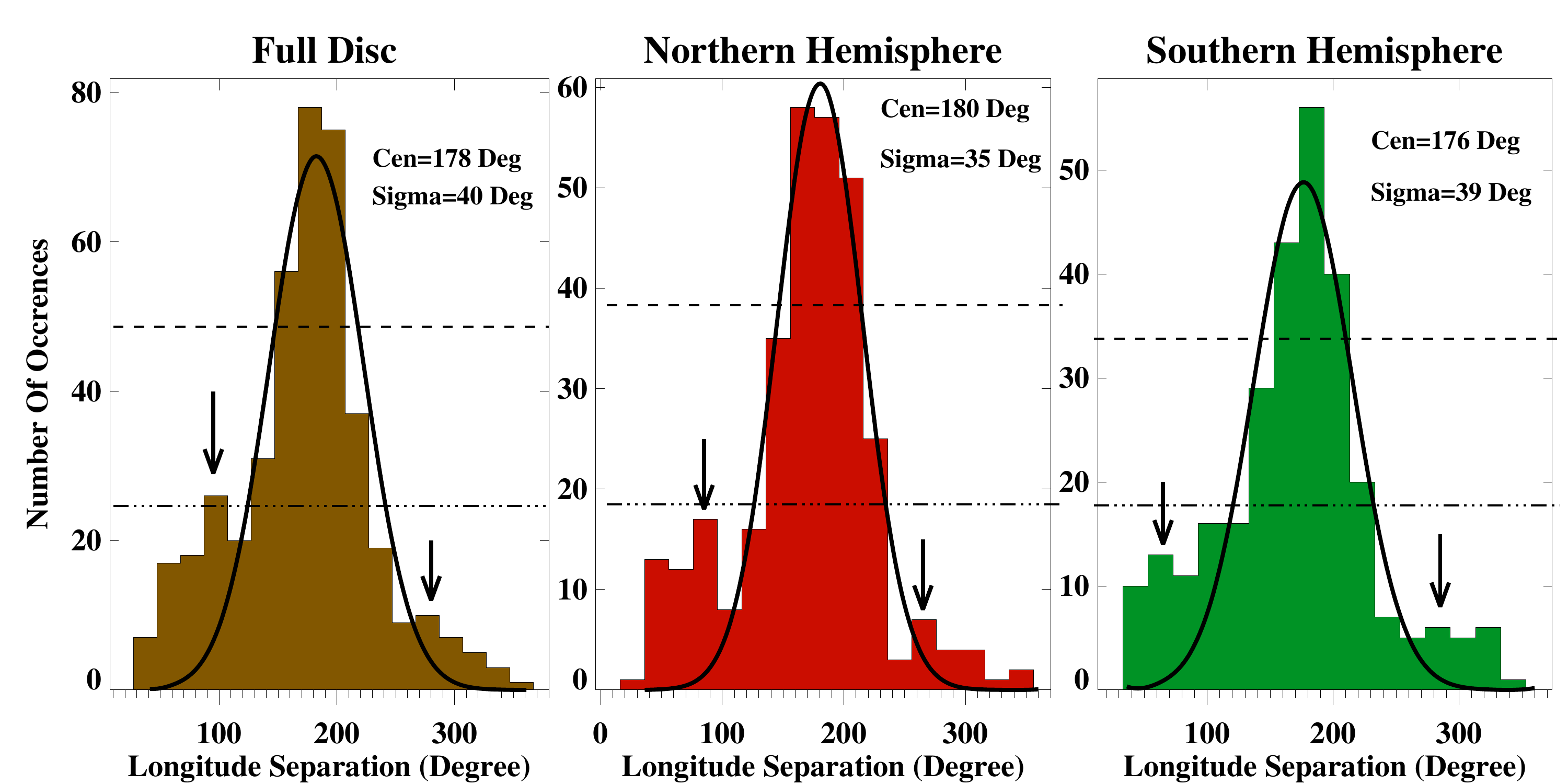}

\caption{Histograms of the longitude separation computed using the `bolometric curve' method. Fitted Gaussian functions to every histograms are shown in thick black line. Fitted Gaussian parameters are printed in corresponding panels. Two black arrows point towards the other two peaks of the histograms located at $\sim$90\textdegree~ and $\sim$270\textdegree~. Two horizontal lines (dash-dot and dash-dash) represent the mean and the (mean+$\sigma$) values respectively.} 

\label{b_histo}
\end{figure*}
Similar to the earlier method, in this case also we have calculated the longitude separation between the two most active spot concentrations for every Carrington map and plotted the histograms as shown in Figure~\ref{b_histo}. We can clearly see that for every case (full disc, northern and southern hemisphere) the histograms peak at $\sim$180\textdegree~. In each case, the histogram distribution look similar to a bell-shaped curve. We thus fit every distribution with a Gaussian function as shown by the solid black lines in Figure~\ref{b_histo}. The centers of the fitted Gaussians, for the three cases, are at 178\textdegree~, 180\textdegree~, 176\textdegree~. This agrees well with the results found by \citet{2003A&A...405.1121B,2005A&A...441..347U}. Apart from the well structured peak at $\sim$180\textdegree~, we also highlight the two other peaks (though considerably weaker) at $\sim$90\textdegree~ and $\sim$270\textdegree~ by two arrows in Figure~\ref{b_histo}. We remind the reader here that the these peaks were also found from the rectangular grid method (Figure~\ref{u_histograms}). These two peaks at $\sim$90\textdegree~ and $\sim$270\textdegree~ probably arise due to the dynamic nature of the active longitude locations. Apart from that, there could also have been some contributions due to the different sunspot sizes on the active longitude separations \citep{2007AdSpR..40..959I}.

\begin{figure*}[!htbp]
\centering
\includegraphics[width=0.85\textwidth]{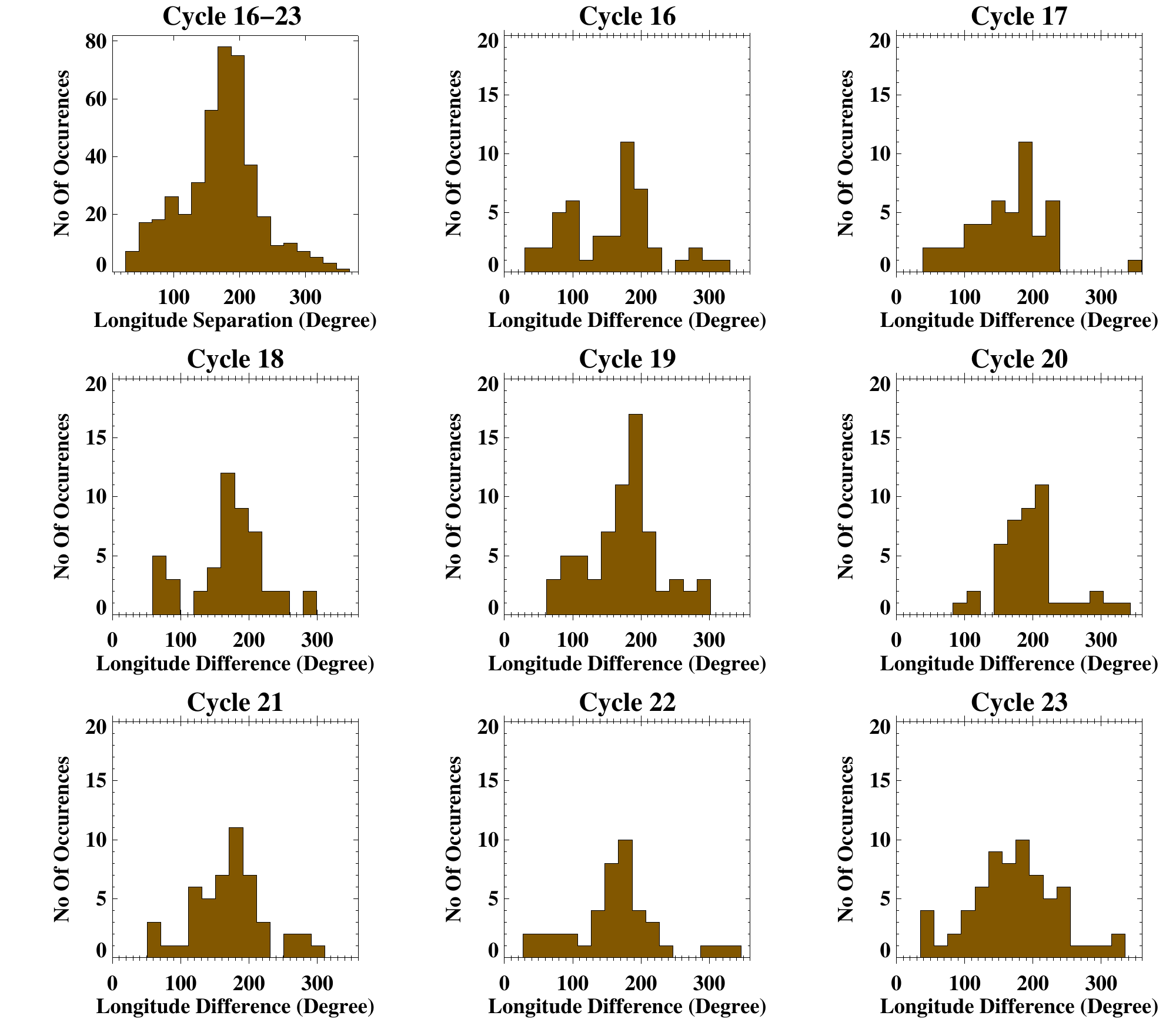}

\caption{Different histograms showing the longitude separation for the solar cycles 16-23. } 

\label{b_cycles}
\end{figure*}

Next we investigate the occurrences of the peaks found in Figure~\ref{b_histo} for every individual solar cycle (cycle 16 to 23). Different panels of Figure~\ref{b_cycles} show the longitude separation histograms for the full period as well as for the individual cycles. We notice that the separation peaks at 180\textdegree~ for every cycle and the height of this peak follow the cycle strengths i.e the strongest cycle, cycle 19 in this case, has the maximum number of occurrences at 180\textdegree~ and so on. Also we notice that the two other peaks (at $\sim$90\textdegree~ and $\sim$270\textdegree~) are also present in most of the cycles, though with lesser strengths. Thus we confirm that these active longitudes persist for the whole 90 years of data analyzed in this paper.

\section{Periodicities In Active Longitudes} \label{period}

Active longitudes have been shown to migrate with the progress of the solar cycle \citep{2005A&A...441..347U}.
\begin{figure*}[!htbp]
\centering
\includegraphics[width=0.97\textwidth]{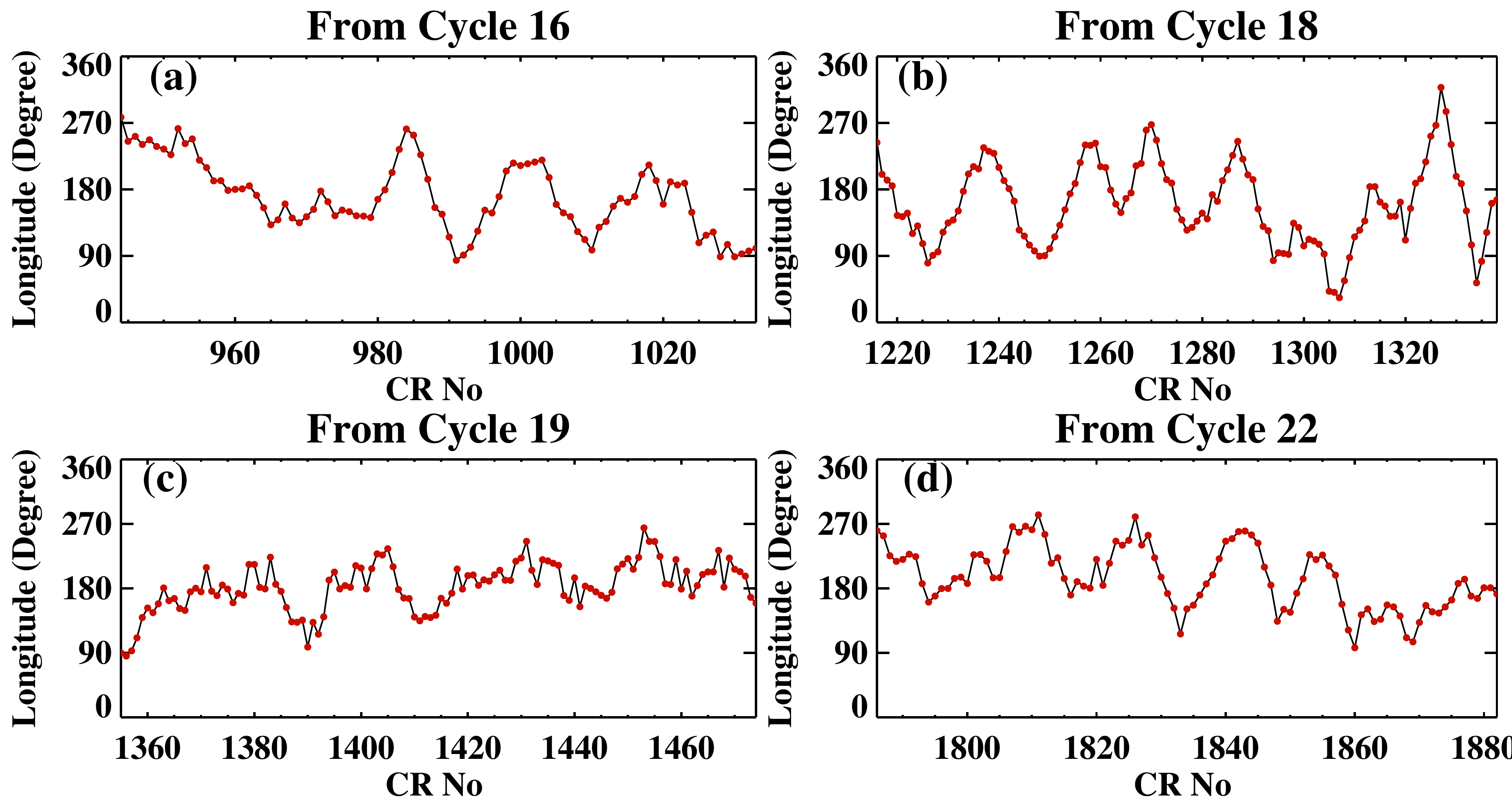}

\caption{ Panels (a-d) show the time variation of the longitude of maximum activity, as found from the method of `bolometric curve', for four different cycles. A running average of 6 months has been performed to suppress the small fluctuations.}  
\label{b_lc}
\end{figure*}
\begin{figure*}[!htbp]
\centering
\includegraphics[trim = 3mm 0mm 105mm 0mm, clip,width=0.30\textwidth,angle=90]{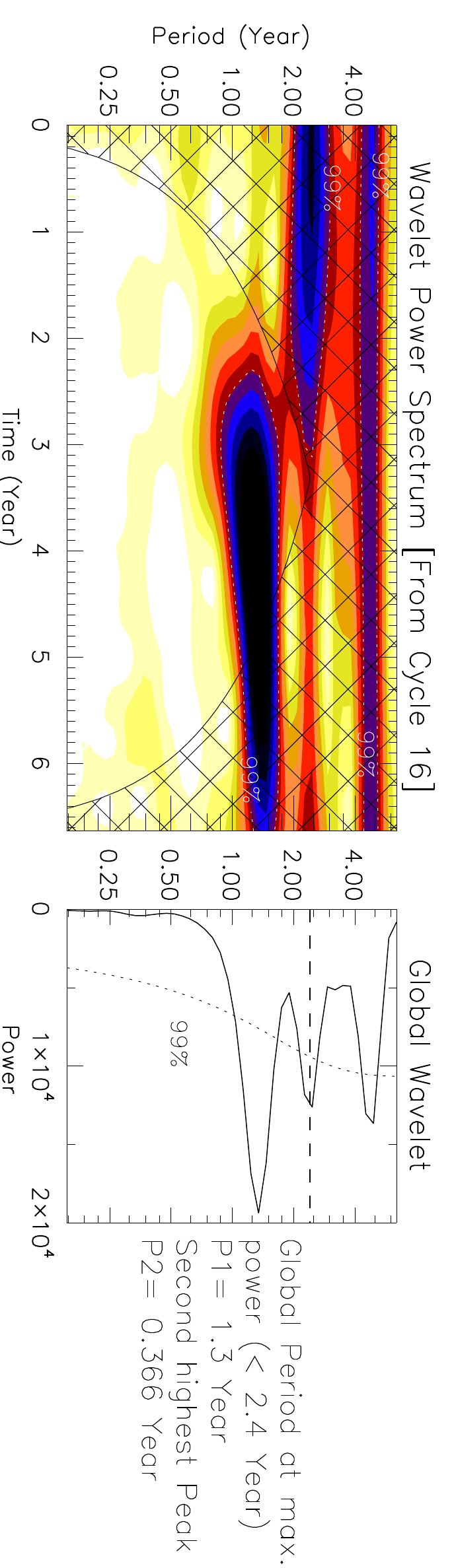}
\includegraphics[trim = 3mm 0mm 105mm 0mm, clip,width=0.30\textwidth,angle=90]{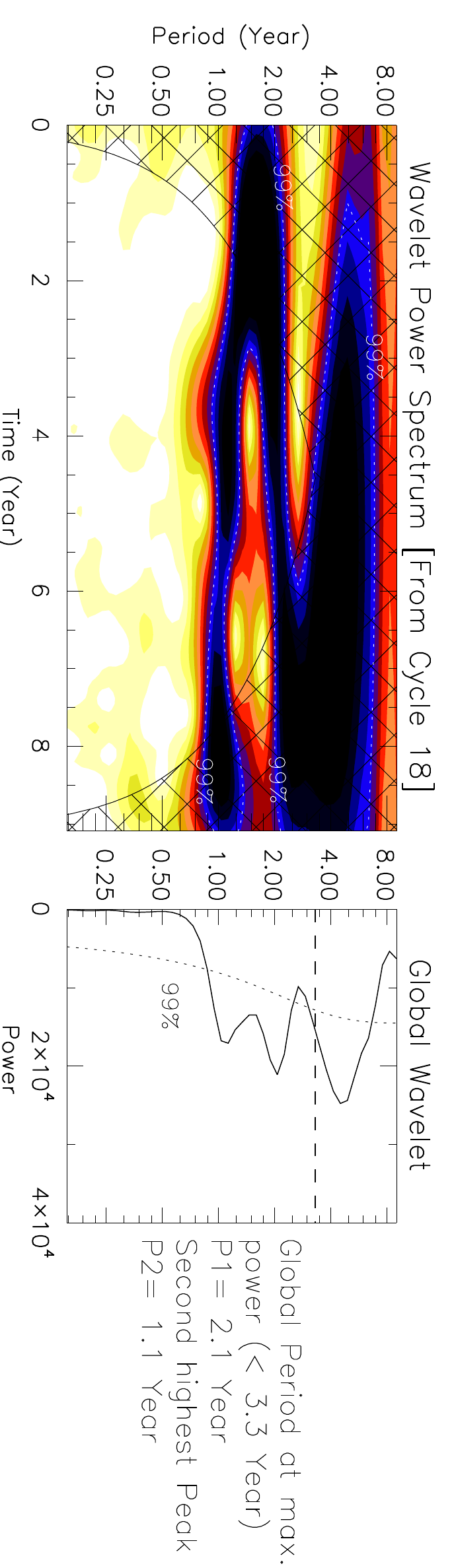}
\includegraphics[trim = 3mm 0mm 105mm 0mm, clip,width=0.30\textwidth,angle=90]{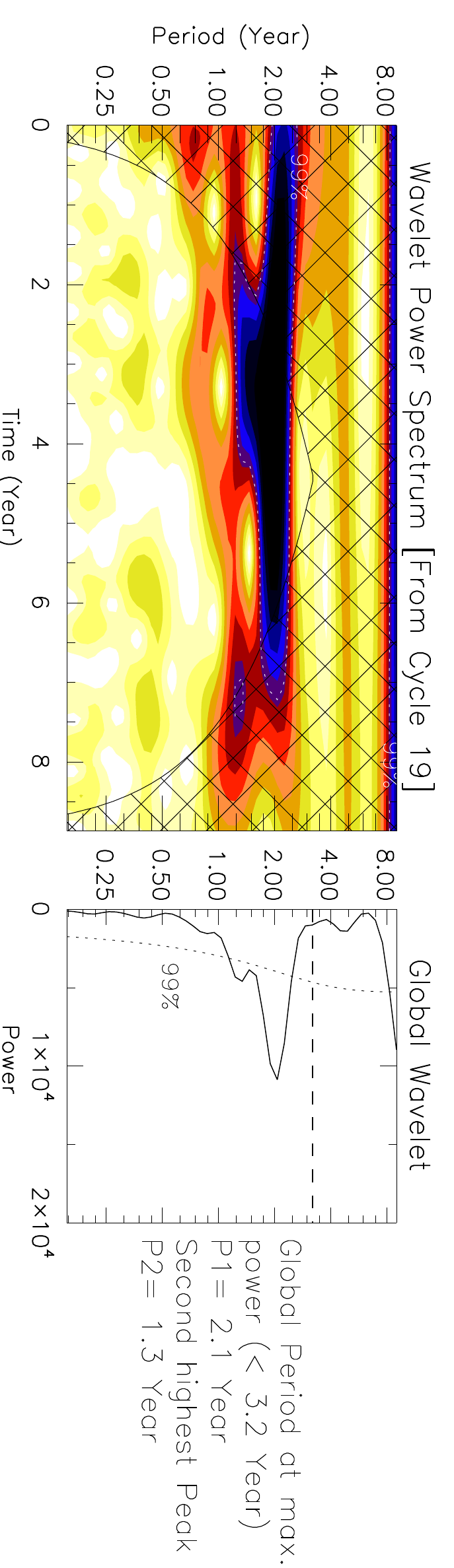}
\includegraphics[trim = 3mm 0mm 105mm 0mm, clip,width=0.30\textwidth,angle=90]{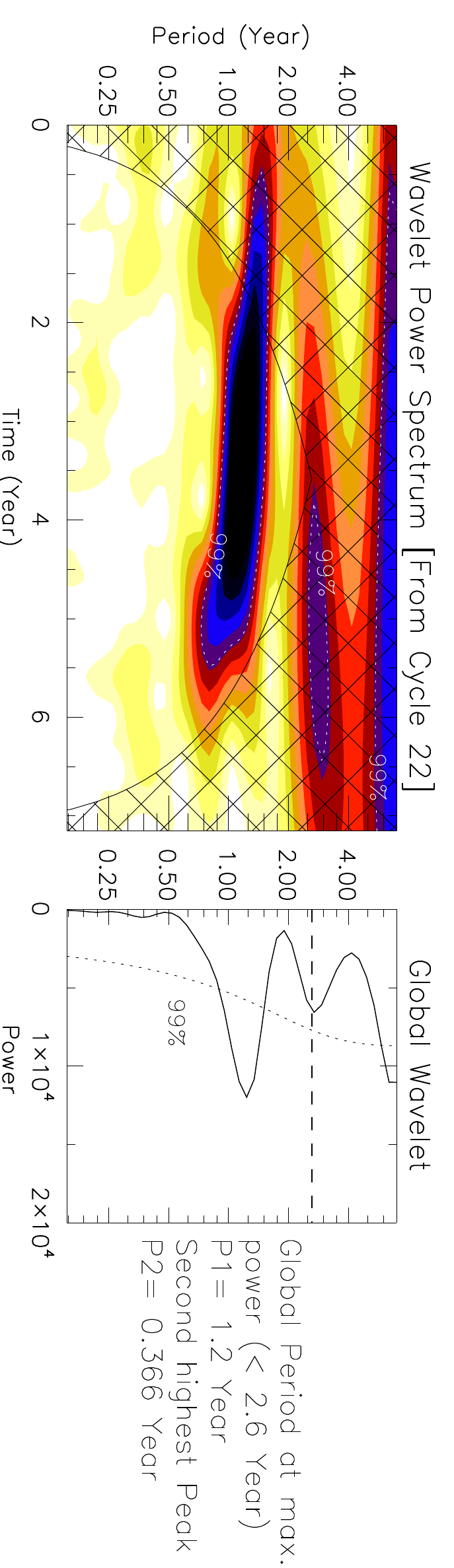}

\caption{[Top to bottom]: Results of wavelet analysis on the light curves shown in panels (a-d) of Figure~\ref{b_lc} respectively. The periods with the maximum significant powers are listed after the wavelet power spectrum (left panel) and the global wavelet plot (middle panel).} 

\label{wave_obs}
\end{figure*}
 Also the the activity switches periodically between the two most active longitude zones \citep{2003A&A...405.1121B}. We investigate the same by using the longitude information of the maximum dip (L$_{m}$) using the `bolometric curve' method (for an example, the 330\textdegree~ longitude of panel (c) of Figure~\ref{b_image}). Since we have rejected any Carrington map which has a data gap (due to missing days in the original Kodaikanal data), we do not have a continuous stretch of L$_{m}$ for more than 8 years.
  Figure~\ref{b_lc} shows the time variation of the L$_{m}$ for four different solar cycles for which we have minimum of 6 years of continuous values of L$_{m}$. To smooth out the small fluctuations, we have performed running averaging of 6 months (following \citealp{2003A&A...405.1121B}). From the plot, we clearly identify periodic variations in every light curves. To get a quantitative estimation of the periods, we use the wavelet tool. Results from the wavelet analysis on the L$_{m}$ light curves ( panels (a-d) of Figure~\ref{b_lc}) are shown in Figure~\ref{wave_obs}. In all these plots, left panel shows the wavelet power spectrum and the right panel shows the global wavelet power which is nothing but the wavelet power at each period scale averaged over the time. The 99\% significance level calculated for the white noise \citep{1998BAMS...79...61T} has been represented by the contours shown in the wavelet plot and by the dotted line plotted in the global wavelet plot. The effect of the edges represented by the cone of influence (COI), has been shown as the cross-hatched region. Obtained periods are indicated in the right hand side of each plots. Here we must highlight the fact that due to the shorter time length of the light curves ($\leq$ 9 years) the maximum measurable period in the wavelet (due to COI) is always $\leq$ 3.5 years. 

  The global wavelet plots indicate two prominent periods of 1.3 years and 2.1 years. This means that the position of the most active bin moves periodically and these periods persist over all cycles investigated in this case. The occurrence of these two periods are particularly interesting as they have been found using the sunspot area time series from different observatories around the world \citep{1990A&A...238..377C,2002A&A...394..701K,2006ChA&A..30..393Z,2016arXiv160804665M}. Also the presence of these periods in all the cycle again confirm their connection with the global behavior of the solar cycle.

We notice in Figure~\ref{b_lc} that there is an average drift of the longitude of maximum activity with the progress of the cycles and this probably is connected with the solar differential rotation as explored in the following section.

  \subsection{Differential Rotation and the phase curve} \label{theory_mig}

Previous studies have shown that the migration of active longitudes is governed by the solar differential rotation. According to \citet{2003A&A...405.1121B,2005A&A...441..347U}, the migration pattern can be easily explained if one uses the differential rotation profile suitably. Thus, we move over to a dynamic reference frame defined by solar differential rotation as described below.

\begin{figure}[!htbp]
\centering
  \includegraphics[width=0.85\linewidth]{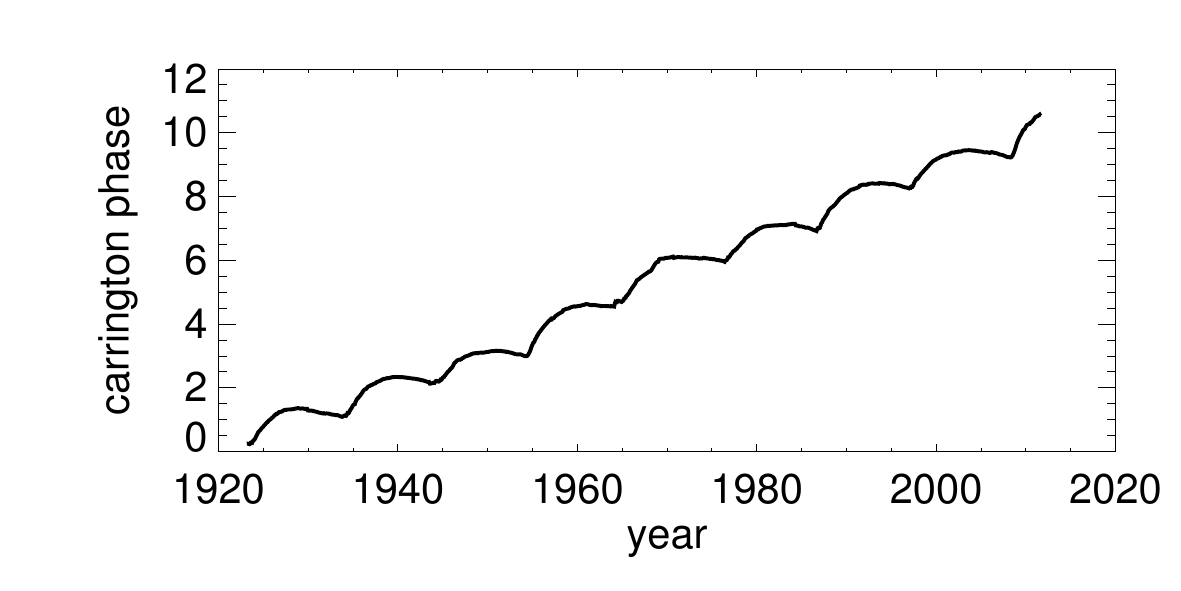}
    \caption{Carrington phase of active longitude as a function of time as derived using the Equation~\ref{theory}.}
  \label{fig:phase}
\end{figure}

  The rotation rate of the longitude of activity for the $i^{th}$ Carrington rotation can be expressed as,
\begin{equation}
  \Omega_{i}=\Omega{_0}-B\sin^2\langle\theta\rangle_i
\end{equation}
 where $\langle\theta\rangle_i$ denotes the sunspot area weighted latitude with ${\Omega}_0$ and $B$ being $14.33^{\circ}$/day and $3.40$ respectively (following \citet{2005A&A...441..347U}). Using this rotation rate the longitudinal position of active longitude in Carrington frame for the $i^{th}$ rotation (${\Lambda}_i$) can be calculated from the same at $N_0^{th}$ rotation (${\Lambda}_0$) through the relation defined as,
\begin{equation}
  \Lambda{_i}=\Lambda{_0}+T_C\sum_{j=N_0+1}^{i} ({\Omega}_C-{\Omega}_j) 
  \label{theory}
\end{equation}
with ${\Omega}_C=\frac{360^{\circ}}{T_C}$ and $T_C=25.38$ days. From the longitudes, phases are calculated as $\phi=\frac{\Lambda}{360^{\circ}}$. These phases are made continuous (Figure~\ref{fig:phase}) by minimizing $\left|\phi_{i+1}+N-\phi_i\right|$ with $N$ spanning over positive and negative integers. $\phi_{i+1}$ is replaced by $\phi_{i+1}+N$ for the $N$ which gives the minimum absolute difference mentioned. We calculate the missing phases to fill the gaps occurred due to missing Carrington maps by interpolating over $\langle\theta\rangle_i$. From Figure~\ref{fig:phase}, we see quite a few distinct features. Immediately we recover the 11-year period of the solar cycle. Also, we see that for an individual cycle the curve first steepens and then dips towards the later half of the cycle. This is explained by considering the fact that in the beginning of a solar cycle, sunspots appear at higher latitudes where the rotation rate is quite different from the Carrington rotation rate (which is basically the rotation rate at $\approx$15\textdegree~ latitude). As the cycle progress, sunspots move down towards the equator and the curves then tends to flatten. 
 \begin{figure}[!htbp]
\centering
  \includegraphics[width=0.85\linewidth]{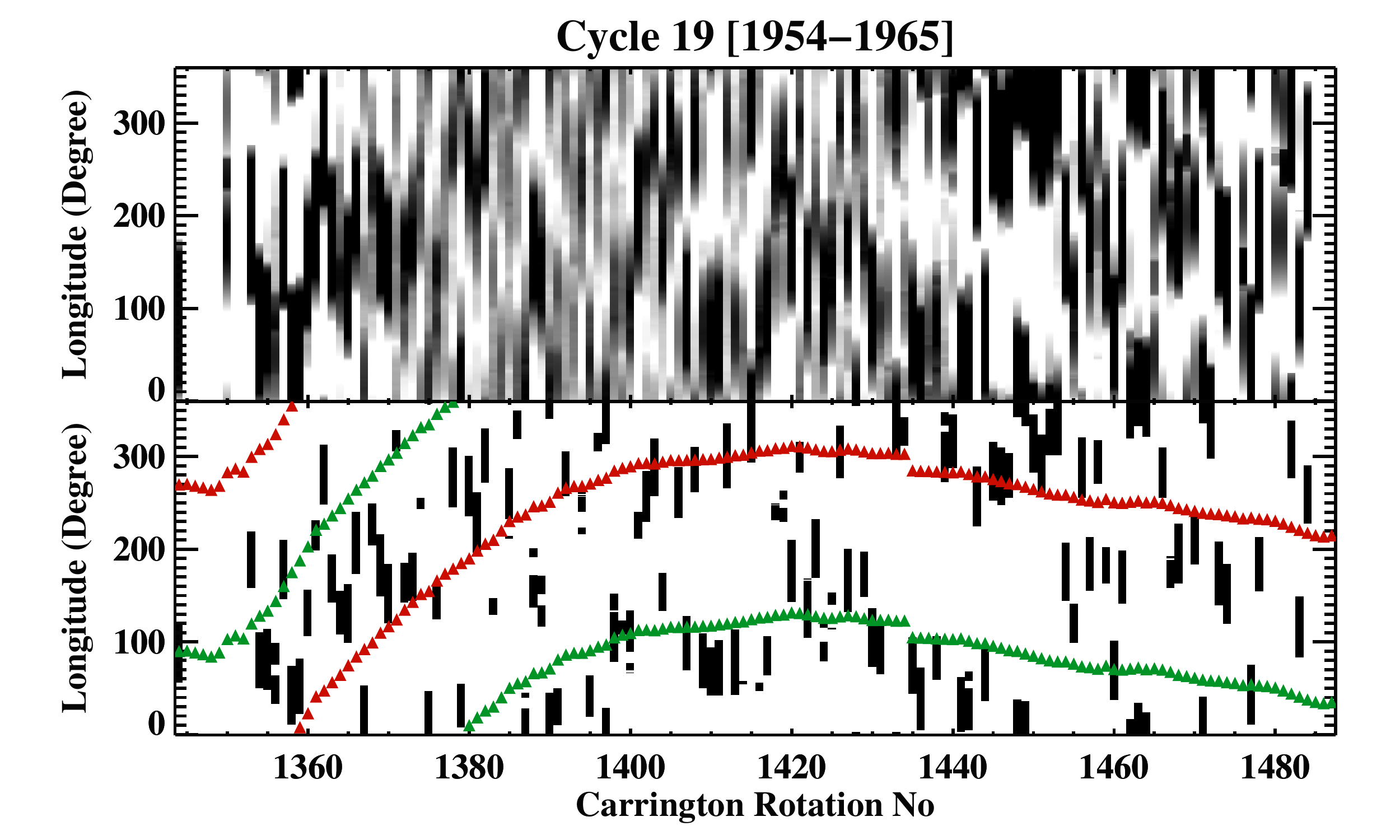}
    \caption{ Top panel shows the bolometric profiles stacked over Carrington rotation period for the 19$^{th}$ cycle. Sigma thresholded image of the same is shown in the bottom panel. Overplotted red and green symbols indicate the phases as obtained using Equation~\ref{theory}.}
  \label{19_cycle}
\end{figure} 

Though we call it a `theoretical curve' but we want to remind the reader that the area weighted latitude information is extracted  for the generated Carrington maps and thus it will be appropriate to call it a `data driven theoretical curve'.

 We use this `theoretical curve' to demonstrate the association of the migration of the active longitudes with the solar differential rotation. In the top panel of Figure~\ref{19_cycle} we plot the full disc bolometric profiles (as obtained previously) and  stacked them over Carrington rotation for the period 1954 to 1965 (corresponds to 19$^{th}$ cycle). The two dark curves are the manifestation of the two dips corresponding to two active longitudes. To highlight this trend more, we use sigma thresholding (i.e mean+$\sigma$) on the original image and plotted it in the bottom panel of Figure~\ref{19_cycle}. We then generate the theoretical curves, corresponding to two active longitudes as obtained from each bolometric profiles and overplotted them. A good match between the theoretical curve and the obtained active longitude positions confirms the fact that the migration of these active longitude is indeed dictated by the solar differential rotation. Here we again highlight the fact that the missing phases have been filled using the interpolation method. Since the current phase has contributions from the previous phases (see Equation~\ref{theory}),  we could not match every details of the observed pattern for all the cycles. The small discrepancy between the theoretical curve and the data could also be due to the fixed values of the differential rotation parameters used in this study which may not be suitable for all cycles.

\section{Summary and Conclusions}

In this paper in the context of active longitude, we have analyzed, for the first time, the Kodaikanal white-light digitized data which covers cycles 16-23. We have analyzed the data with two previously known methods: `rectangular grid' method and the `bolometric curve' method for the full disc as well as for the individual hemispheres. Below we summarize the key findings from the two methods:

$\bullet$ From the two methods, we see that for the entire duration of the data analyzed, we find two persistent longitude zones or `active longitudes' with higher activity. This is consistent with the results from the Greenwich data as obtained by \citet{2003A&A...405.1121B}.

$\bullet$ Using the `rectangular grid' method we have constructed the histograms of the longitude separation between the two active longitudes and found prominent peaks in the histograms at $\sim$90\textdegree~, $\sim$180\textdegree~ and at $\sim$270\textdegree~. We also noted that the highest peak occurs for the separation of 20\textdegree~. We use area thresholding on the sunspots to show that the peak at 20\textdegree~ is due to the presence of relatively large sunspots being shared by two consecutive longitude bins. 

$\bullet$ Using the bolometric method we recover the peaks at $\sim$90\textdegree~, $\sim$180\textdegree~ and at $\sim$270\textdegree~ as found earlier. Also, we found that the peak height for the 180\textdegree~ separation is much higher than the other two peaks. We fitted the central lobe with a Gaussian function and estimated the center location. Applying this method for individual solar cycles we established that the peak at 180\textdegree~ is always present in every solar cycle.

$\bullet$ Using temporal evolution of the peak location of highest activity we have demonstrated the presence of two periods using the wavelet analysis. The two prominent periods are 1.1-1.3 years and 2.1-2.3 years. These two periods are routinely found in the sunspot area and sunspot number time series. Apart from that we also observe another period of $\sim$5 years with significant power. Due to shorter length of the time series, this period is beyond the detection confidence level. However the presence of the period directly indicate its connection with the global solar dynamo mechanism which needs to be investigated further. 

$\bullet$ Finally , we use the solar differential rotation profile to construct a dynamic reference frame. A theoretical curve has been generated using area weighted sunspot latitude information from the Carrington maps. While overplotting this curve on top of the sigma thresholded image of the bolometric profiles, we have shown that the migration pattern follows the solar differential rotation as found in some of the previous studies.

To conclude, we found signatures of persistent active longitudes on the Sun using the Kodaikanal data. We hope that with these observational results along with the solar models, understanding of the physical origin of active longitudes can be advanced.

\section{Acknowledgment}
The authors would like to thank Mr. Gopal Hazra for his useful comments in preparing this manuscript. We thank the reviewer for his/her constructive comments and suggestions which improved the content and presentation of the paper. We would also like to thank the Kodaikanal facility of Indian Institute of Astrophysics, Bangalore, India for proving the data. This data is now available for public use at \url{http://kso.iiap. res.in/data}.



 \bibliographystyle{apj}

\end{document}